\documentclass[aps,prl,twocolumn,showpacs]{revtex4}
\usepackage{amsmath}
\usepackage{amssymb}
\newcommand\tr{\mathbf{tr}}
\newcommand\di{\mathrm{d}}
\newcommand\half{{\frac{1}{2}}}
\newcommand\quar{{\frac{1}{4}}}
\newcommand\oct{{\frac{1}{8}}}
\newcommand\ro{{\hat{\rho}}}
\newcommand\Ao{\hat{A}}
\newcommand\Bo{\hat{B}}
\newcommand\Co{\hat{C}}
\newcommand\qo{\hat{q}}
\newcommand\po{\hat{p}}
\newcommand\Pio{\hat{\Pi}}
\newcommand\sio{\hat{\sigma}}
\newcommand\si{\sigma}
\newcommand\Po{\hat{P}}
\newcommand\Mb{\mathbf{M}}
\newcommand\ev{\vec{e}}
\newcommand\keti{\vert i\rangle}
\newcommand\ketf{\vert f\rangle}
\newcommand\brai{\langle i\vert}
\newcommand\braf{\langle f\vert}
\newcommand\ketup{\vert\!\uparrow\rangle}
\newcommand\ketdn{\vert\!\downarrow\rangle}
\newcommand\braup{\langle\uparrow\!\vert}
\newcommand\bradn{\langle\downarrow\!\vert}
\begin{document}
\title{Structural Features of Sequential Weak Measurements}    
\author{Lajos Di\'osi}
\email{diosi.lajos@wigner.mta.hu}
\affiliation{
Wigner Research Centre for Physics\\
H-1525 Budapest 114, POB 49, Hungary}
\date{\today}

\begin{abstract}
We discuss the abstract structure of sequential weak measurement (WM) of general observables.
In all orders, the sequential WM correlations without post-selection yield the corresponding correlations 
of the Wigner function, offering direct quantum tomography through the moments of the canonical variables.
Correlations in spin-$\half$ sequential weak measurements coincide with those in strong measurements, 
they are constrained kinematically, are equivalent with single measurements. 
In sequential WMs with post-selection, a new anomaly occurs, 
different from the weak value anomaly of single WMs. In particular, the spread of polarization $\sio$, as 
measured in double WM of $\sio$, will diverge for  certain orthogonal pre- and post-selected states.  
\end{abstract}

\pacs {03.65.Ta, 03.65.Wj}

\maketitle
From textbooks on quantum mechanics we learn that the ideal measurement of 
observable $\Ao$ collapses the pre-measurement state $\ro$ into an eigenstate of  $\Ao$ 
hence erasing all memory of $\ro$. If the measurement is non-ideal (i.e.: unsharp, imprecise)  the collapse 
still happens although it may keep some well-defined features of $\ro$.
On one hand, the larger the unsharpness the more faithfully the pre-measurement state will be preserved.
On the other hand, the imprecision of the measurement can be compensated by measuring on
a larger ensemble of identically prepared pre-measurement states. The concept of weak measurement (WM)  
corresponds to the asymptotic limit of zero precision and infinite statistics \cite{Dio06} when
the pre-measurement state $\ro$ would invariably survive the measurement. 
WM was used by Aharonov, Albert and Vaidman \cite{AAV88}
as a non-invasive quantum measurement between pre-selection (preparation) and post-selection of the pre- and
post-measurement states, respectively. Non-invasiveness of WM is a remarkable feature both with and without
post-selection, and this non-invasiveness can be maintained for a succession of WMs on a single quantum
system. General features of such sequential WMs form the subject of the present work.  
  
\emph{WMs without post-selection} --- We outline WM of a single observable $\Ao$ at the abstract level of generalized (unsharp, imprecise)
measurements \cite{POVM}. Consider the pre-measurement state $\ro$ and the unsharp measurement of $\Ao$, with precision $a$.
Let $G_a(A)$ stand for a Gaussian function of standard width $a$. The unnormalised post-measurement state 
conditioned on the outcome $A$ takes this form:
\begin{equation}
\label{ro_A}
\ro_a(A)=\sqrt{G_a (A-\Ao)}\ro\sqrt{G_a (A-\Ao)}
\end{equation}
where the outcome probability  satisfies
\begin{equation}
\label{p_A}
p_a(A)= \tr\ro_a(A)= \langle G_a(A-\Ao) \rangle_\ro\;.
\end{equation}
If we calculate the stochastic mean $\Mb A$  of $A$ we get
\begin{equation}
\label{M_A}
\Mb A = \int p_a(A) A \di A= \langle\Ao\rangle_\ro\;.
\end{equation}
We are interested in the WM limit of infinite imprecision $a\rightarrow\infty$, i.e., when  
$a$ is so large that the  difference between pre- and post-measurement states is negligible.
In practice it means $a\gg\Delta A$ where $(\Delta A)^2=\langle\Ao^2\rangle_\ro-(\langle\Ao\rangle_\ro)^2$.
While the relationship $\Mb A$ is independent of $a$  the probability distribution $p_a(A)$ diverges  so that
$p_\infty(A)$ does not exist. Note with ref. \cite{Dio06} the WM limit of the unsharp measurement (\ref{ro_A},\ref{p_A}) 
had been used earlier for theory of time-continuous measurement \cite{contmeas}.
 
Before constructing sequential WMs, let us write the post-measurement state \eqref{ro_A} into the equivalent form:
\begin{equation}
\label{ro_A_supop}
\ro_a(A)=\exp\left(\frac{-\Ao_\Delta^2}{8a^2}\right)G_a (A-\Ao_c)\ro
\end{equation}
where $\Ao_\Delta,\Ao_c$ are commuting superoperators \cite{supop} defined by 
$\Ao_\Delta\hat{O} =[\Ao,\hat{O}]$ and $\Ao_c\hat{O}=\half\{\Ao,\hat{O}\}$.
As an example of sequential WMs, we consider the sequence of three
independent WMs of $\Ao,\Bo,\Co$, in this order. In the WM limit, we can apply eq. \eqref{ro_A_supop} without the
exponential factor to construct the unnormalized post-measurement state:
\begin{equation}
\label{ro_ABC}
\ro_a(A,B,C)=G_a (C-\Co_c)G_a (B-\Bo_c)G_a (A-\Ao_c)\ro.
\end{equation}
The joint probability distribution of the three outcomes is determined by the trace of the post-measurement state:
\begin{eqnarray}
\label{p_ABC}
&&p_a(A,B,C)=\tr\ro_a(A,B,C)=\\
                       &=&\tr\bigl\{G_a (C-\Co_c)G_a (B-\Bo_c)G_a (A-\Ao_c)\ro\bigr\},\nonumber
\end{eqnarray}
which, as we said already, diverges in the WM limit and $p_\infty(A,B,C)$ does not exist. Nonetheless, 
the stochastic average of the product ABC is independent of $a$ in the WM limit. Using eq. \eqref{p_ABC}, we obtain
\begin{eqnarray}
\label{M_ABC}
\Mb ABC &=& \int p_a(A,B,C) ABC~ \di A \di B \di C=\nonumber\\
                  &=&\frac{1}{8}\bigl\langle\{\Ao,\{\Bo,\Co\}\}\bigr\rangle_\ro\;.
\end{eqnarray}
This important result was obtained  by Bednorz and Belzig \cite{BedBel10} assuming a quasi-distribution which this time
we justify as follows.

Since the  r.h.s. of the above expression is independent of $a$ therefore we can calculate it for $a=0$.
This means, we get the following quasi-distribution from the true $p_a(A,B,C)$:
\begin{equation}
\label{p_ABC_quasi}
p_0(A,B,C)=\tr\bigl\{\delta(C-\Co_c)\delta(B-\Bo_c)\delta(A-\Ao_c)\ro\bigr\}.
\end{equation}
 This quasi-distribution can have negative domains. (For the true distribution $p_a(A,B,C)\geq0$ holds in the WM limit.) 
 The merit of this quasi-distribution is that it does not contain the diverging parameter $a$ and yields correctly the
 mean for the product $ABC$ exactly like $p_a(A,B,C)$ did:
 \begin{eqnarray}
\label{M_ABC_quasi}
\Mb ABC &=& \frac{1}{8}\bigl\langle\{\Ao,\{\Bo,\Co\}\}\bigr\rangle_\ro=\nonumber\\
                  &=&\int p_0(A,B,C) ABC~ \di A \di B \di C.
\end{eqnarray}
The same is true for the means of $A,B,C,AB,AC,BC$, respectively.
But all other means diverge in reality,  i.e.: with $p_a(A,B,C)$ in the WM limit, whereas $p_0(ABC)$ suggests
incorrect finite values for them. 

The above results can trivially be extended for an arbitrary long sequence of WMs. Let us consider a sequence of
observables $\Ao_1,\Ao_2,\dots,\Ao_n$ which are weakly measured in the given order on the pre-measurement state $\ro$.
If $A_1,A_2,\dots,A_n$ denote the corresponding measurement outcomes then
\begin{equation}
\label{M_As}
\Mb A_1 A_2\dots A_n = \frac{1}{2^n}\bigl\langle\{\Ao_1,\{\Ao_2,\{\dots,\{\Ao_{n-1},\Ao_n\}\dots\}\}\}\bigr\rangle_\ro\;.
\end{equation}
The stochastic mean of the product of sequential WM outcomes coincides with the quantum expectation value
of the \emph{stepwise-symmetrized} (also called time-symmetric \cite{Bergetal09}) product of quantum observables.
This is the central result for sequential WMs without post-selection.
If we regard a subset of the $n$ outcomes and discard the rest of them then the similar identity holds, e.g.:
$\Mb A_2 A_7 A_8 = \frac{1}{8}\bigl\langle\{\Ao_2,\{\Ao_7,\Ao_8\}\}\bigr\rangle_\ro$.
In general,
\begin{equation}
\label{M_As_subset}
\Mb A_{i_1}\dots A_{i_2}\dots A_{i_r} = \frac{1}{2^r}\bigl\langle\{\Ao_{i_1},\{\Ao_{i_2},\dots,\Ao_{i_r}\}\}\bigr\rangle_\ro\;
\end{equation}
holds for $(i_1,i_2,\dots,i_r)\subseteq(1,2,\dots,n)$, i.e., for any ordered subsets of indices from $1$ to $n$,
as it follows easily from the derivation of  eq. \eqref{M_As}. Also we can derive all the above stochastic means
from the quasi-distribution:
\begin{eqnarray}
\label{p_As_quasi}
& &p_0(A_1,A_2,\dots,A_n)=\\
&=&\tr\bigl\{\delta(A_n-\Ao_{1,c})\delta(A_2-\Ao_{2,c})\dots\delta(A_n-\Ao_{n,c})\ro\bigr\}.\nonumber
\end{eqnarray}
Generalization of the relationship \eqref{M_ABC_quasi} holds:
 \begin{eqnarray}
\label{M_A'_An_quasi}
& &\Mb A_1 A_2\dots A_n = \frac{1}{2^n}\bigl\langle\{\Ao_1,\{\dots,\{\Ao_{n-1},\Ao_n\}\dots\}\}\bigr\rangle_\ro=\nonumber\\
                                          &=&\int p_0(A_1,A_2,\dots,A_n) A_1 A_2\dots A_n~ \di A_1 \di A_2 \dots \di A_n.
\end{eqnarray}

The last two WMs in a stepwise-symmetrized sequence are always interchangeable but the rest of them are not: order of WMs matters in general.
There is, however, a remarkable class when all WMs are interchangeable. Let us discuss the example of the
sequence $\Ao,\Bo,\Co$. To find a simplest sufficient condition of complete interchangeability, we require that the superoperators $\Ao_c,\Bo_c,\Co_c$ in eq. \eqref{M_ABC} all commute.  Take, e.g.,  the identity  $[\Ao_c,\Bo_c]\hat{O} =\frac{1}{4}[[\Ao,\Bo],\hat{O}]$ which says that  $\Ao_c$ and $\Bo_c$
commute if $[\Ao,\Bo]$ is a c-number. Therefore the interchangeability of the three WMs is ensured if all three
commutators $[\Ao,\Bo]$, $[\Ao,\Co]$ and $[\Bo,\Co]$ are c-numbers.
In the general case, the order of WMs within the sequence $\Ao_1,\Ao_2,\dots,\Ao_n$ becomes
irrelevant if 
\begin{equation}
\label{commute}
[\Ao_k,\Ao_l]=\mbox{c-number}~~~~~(k,l=1,2,\dots,n).
\end{equation}
This is not necessary just a sufficient condition of complete interchangeability of the $n$ WMs.
Under this condition, the stepwise-symmetrization on the r.h.s. of eq. \eqref{M_As} reduces to symmetrization: 
\begin{equation}
\label{M_As_symm}
\Mb A_1 A_2 \dots A_n = \bigl\langle\mathcal{S}\Ao_1\Ao_2\dots\Ao_n\bigr\rangle_\ro\;,
\end{equation}
where $\mathcal{S}$ stands for symmetrization of the operator product behind it.

\emph{Canonical observables} --- The conditions \eqref{commute} hold typically for the linear combination of canonical variables, e.g., for the choice
\begin{equation}
 \Ao_k=u_k\qo+v_k\po~~~~~~(k=1,2,\dots,n),
\end{equation}
where $[\qo,\po]=i$.
Then symmetrization $\mathcal{S}$ is nothing else than Weyl ordering.
Since the Weyl-ordered correlation functions of canonical variables $\qo,\po$, or of their linear combinations like on r.h.s. of eq. \eqref{M_As_symm}  coincide
with the corresponding correlation functions (moments) calculated from the Wigner  function $W(q,p)$ of $\ro$,
we conclude that the r.h.s. can be re-written in terms of Wigner function correlations:
\begin{eqnarray}
\label{M_M_W}
\Mb A_1 A_2 \dots A_n &=& \int W(q,p) A_1 A_2 \dots A_n \di q \di p\equiv\nonumber\\
                                             &\equiv &\langle A_1 A_2 \dots A_n \rangle_W.
\end{eqnarray}
This means that for sequential WMs of canonical observables the generic quasi-distribution \eqref{p_As_quasi} is redundant
for $n>2$,  its role is taken over by the Wigner quasi-distribution. The coincidence  $p_0(q,p)=W(q,p)$ in the special case $n=2,\Ao_1=\qo,\Ao_2=\po$ 
was recognized in \cite{BedBel10}.

Suppose, for instance, we perform two WMs of $\qo$ with outcomes $q_1,q_2$ and two WMs of $\po$ with
outcomes $p_1,p_2$. Then independently of the orders of the four WMs, a sufficiently large
statistics of outcomes allows us to determine all second order moments of the Wigner function
\begin{eqnarray}
\label{qp_moments}
&&\langle q^2\rangle_W= \Mb q_1 q_2,~~~
\langle p^2\rangle_W= \Mb p_1 p_2,\\
&&\langle qp\rangle_W= \Mb q_1 p_1= \Mb q_1 p_2= \Mb q_2 p_1= \Mb q_2 p_2\nonumber
\end{eqnarray}
as well as a few higher order ones $\langle q^2 p\rangle_W$, $\langle qp^2\rangle_W$, $\langle q^2 p^2\rangle_W$ and, of course, 
the first order moments $\langle q\rangle_W$, $\langle p\rangle_W$, too.

\emph{Spin-$\half$ observables} --- Sequential measurement of spin-$\half$ observables is exceptional: 
eq. \eqref{M_As} is valid no matter the measurements are weak, ideal (strong),
or even alternating within the sequence between the two extreme strength. 
Consider the following choice of  observables:
\begin{equation}
\label{sigma_k}
\Ao_1=\sio_1,~\Ao_2=\sio_2,~\dots,\Ao_n=\sio_n
\end{equation}
where $\sio_k$ is the polarization parallel to the unit vector $\ev_k$ for $k=1,2,\dots,n$. 
Denote the measurement  outcomes by $A_1=\si_1,A_2=\si_2$, etc., and invoke eq. \eqref{M_As} form them:
\begin{equation}
\label{M_sigmas}
\Mb \si_1 \si_2\dots \si_n = \frac{1}{2^n}\bigl\langle\{\sio_1,\{\sio_2,\{\dots,\{\sio_{n-1},\sio_n\}\dots\}\}\}\bigr\rangle_\ro\;.
\end{equation}  
To confirm it for strong measurements as well, we introduce the projectors $\Po_\pm=\half(1\pm\sio)$ diagonalizing the Pauli polarization matrix $\sio$. 
Standard expression for sequential strong measurements reads:
\begin{widetext}
\begin{equation}
\Mb \si_1 \si_2\dots \si_n =
\!\!\!\!\!=\tr\!\!\!\sum_{\si_n=\pm1}\!\!\!\!\si_n\Po_{\si_n}^{(n)}\!\dots\!\left(\sum_{\si_2=\pm1}\!\!\!\!\si_2\Po_{\si_2}^{(2)}\!\!\left(\sum_ {\si_1=\pm1}\!\!\!\!\si_1 \Po_{\si_1}^{(1)}\ro\Po_{\si_1}^{(1)}\!\!\right)\!\!\Po_{\si_2}^{(2)} \!\!\right)\!\dots\Po_{\si_n}^{(n)}\nonumber
\end{equation}
\end{widetext}
Observe  the  identity $\sum_{\si=\pm}\si\Po_a\hat{O}\Po_a=\half\{\sio,\hat{O}\}$
valid for auxiliary $2\times2$ matrices $\hat{O}$, apply it $n$-times. We obtain eq. \eqref{M_sigmas}. 
Evaluating its r.h.s. yields
\begin{equation}
\Mb \si_1 \si_2\dots \si_n = \left\{
            \begin{array}{rl}(\ev_1\ev_2) (\ev_3\ev_4)\dots (\ev_{n-1}\ev_n)~~~~&n~\mbox{even}\\
                                          \langle\sio_1\rangle_\ro(\ev_2\ev_3)\dots (\ev_{n-1}\ev_n)~~~~&n~\mbox{odd}
            \end{array}                                                    
\right.
\end{equation}

Outcome correlations of $n$ sequential WMs on a spin-$\half$ system coincide exactly
with the correlations obtained from strong measurements of the same sequence. 
Correlations are kinematically constrained by the chosen directions of polarization measurements. 
For $n$ even, correlations are completely determined by the single mean $\langle\sio_1\rangle_\ro$ 
and just independent of the pre-measurement state $\ro$ if $n$ is even.
 
\emph{WMs with post-selection} ---
So far we have established the general features of outcome statistics  in sequential WMs without
post-selection.  Including post-selection requires straightforward modifications. For mixed state
post-selection \cite{Dio06,Siletal14}, the statistics \eqref{p_ABC} of the $ABC$-sequential WM modifies like this:
\begin{eqnarray}
\label{p_ABC_post}
&&p_a(A,B,C\vert\Pio)= \frac{\tr\bigl\{\Pio\ro_a(A,B,C)\bigr\}}{\tr\bigl\{\Pio\ro\bigr\}}=\\
                       &=&\frac{\tr\bigl\{\Pio G_a (C-\Co_c)G_a (B-\Bo_c)G_a (A-\Ao_c)\ro\bigr\}}{\tr\bigl\{\Pio\ro\bigr\}}\nonumber,
\end{eqnarray}
where $0\leq\Pio\leq1$. Accordingly, the post-selected mean \eqref{M_ABC}, i.e., the mean  restricted for the post-selected 
subset $ABC\vert_{psel}$ of $ABC$, becomes 
\begin{equation}
\Mb ABC\vert_{psel}=\frac{1}{8}\bigl\langle\{\Ao,\{\Bo,\{\Co,\Pio\}\}\}\bigr\rangle_\ro/\bigl\langle\Pio\bigr\rangle_\ro\;.
\end{equation}
The general result must be the following:
\begin{equation}
\label{M_As_post}
\Mb A_1,A_2,\dots,A_n\vert_{psel}
                       =\frac{\bigl\langle\{\Ao_1,\{\Ao_2,\dots,\{\Ao_n,\Pio\}\dots\}\}\bigr\rangle_\ro}{2^n\bigl\langle\Pio\bigr\rangle_\ro}.
\end{equation}
In the basic case, both initial and post-selected states are pure states and we are going to take this option:
$\ro=\keti\brai,\Pio=\ketf\braf$. Then, following Mitchison, Jozsa and Popescu, we introduce the sequential weak values
\begin{equation}
\label{VV_seq}
(A_1,A_2,\dots,A_n)_w=\frac{\bigl\braf\Ao_n\Ao_{n-1}\dots\Ao_1\keti}{\braf i\rangle},
\end{equation}
and re-write eq. \eqref{M_As_post} in time-symmetric form  \cite{Mitetal07}:
\begin{eqnarray}
\label{VV_seq_symm}
&&\Mb A_1,A_2,\dots,A_n\vert_{psel}=\\
&=&\frac{1}{2^n}\sum(A_{i_1},A_{i_2},\dots,A_{i_r})_w (A_{j_1},A_{j_2},\dots,A_{j_{n-r}})_w^\star\nonumber 
\end{eqnarray}
where summation is understood for all partitions $(i_1,i_2,\dots,i_r)\cup(j_1,j_2,\dots, j_{n-r})=(1,2,\dots,n)$
where $i$'s and $j$'s remain ordered. Degenerate partitions  $r=0,n$, too,  must be counted.   
Certain options of reduction, shown above for sequential WMs of canonical or spin-$\half$ observables, may still survive
post-selection, here we are not going to discuss them. We show a particular anomaly, not present in
single post-selected  WM but in sequential WMs, even for simplest ones. 
  
\emph{Re-selection} --- 
Consider the special case $\keti=\ketf$ of post-selection, call it \emph{re-selection}. In the case of a single WM, re-selection is equivalent
with no post-selection:
\begin{equation}
\Mb A = \Mb A\vert_{rsel}=\langle\Ao\rangle_\ro\;. 
\end{equation}
Since WMs are considered non-invasive, we expect that the post-measurement state does not differ
from the initial state $\keti$ in the WM limit, re-selection rate tends to $1$ hence the discarded outcomes would not alter the statistics. 
No doubt, this is the case for a single WM. As to sequential WMs,
however, a glance at \eqref{VV_seq_symm} shows that re-selection does not yield equivalent results with no post-selection \eqref{M_As}.
Even the simplest sequential WM will illustrate the anomaly. 
We consider two WMs, moreover, we consider the case when $\Ao_1=\Ao_2=\Ao$, 
i.e, we weakly measure $\Ao$ twice in a row, yielding outcomes $A_1$ and $A_2$, respectively. Without post-selection,
eq. \eqref{M_As} and with re-selection eq.  \eqref{VV_seq_symm} yield, respectively:
\begin{eqnarray}
\Mb A_1 A_2 &=& \langle i\vert \Ao^2\vert i\rangle,\\
\Mb A_1 A_2\vert_{rsel} &=& \half\langle i\vert \Ao^2\vert i\rangle +\half(\langle i\vert \Ao\vert i\rangle)^2.
\end{eqnarray}
Re-selection decreases $\Mb A_1 A_2$ by half of the squared quantum uncertainty $(\Delta A)^2$
in state $\keti$:
\begin{equation}
\Mb A_1 A_2-\Mb A_1 A_2\vert_{rsel} = \half(\Delta A)^2.
\end{equation}
This is an unexpected anomaly. The reason must lie in the contribution of outcomes \emph{discarded} by re-selection,
i.e.: $\Mb A_1 A_2\vert_{disc}\times\mbox{(discard rate)}\rightarrow\half (\Delta A)^2$ must be satisfied.
 
As an example, consider a spin-$\half$ system in upward polarized initial state $\keti=\ketup$.
Let us begin with a single WM of $\sio\equiv\sio_x$ with outcome $\si_1$. 
The contribution of the discarded outcomes reads
\begin{equation}
\Mb \si_1\vert_{disc} = \frac{\bradn\left(\exp(-\oct\sio_\Delta^2/a^2)\sio_c\ketup\braup\right)\ketdn}
                                                 {\bradn\left(\exp(-\oct\sio_\Delta^2/a^2)\ketup\braup\right)\ketdn}
\end{equation}
where we use the exact expression of the post-WM state with the exponential factor as in eq. \eqref{ro_A_supop} 
otherwise we get $0$ for the rate of discarded events. This rate is
just the denominator in the above fraction, yielding $\sim\quar a^{-2}$ asymptotically.
This rate goes to zero in the WM limit but $\Mb \si_1\vert_{disc}$ vanishes anyway since the numerator is zero
identically. Now, let us weakly measure $\sio\equiv\sio_x$ twice in a sequence, yielding outcomes $\si_1,\si_2$. 
Since the quantum spread $\Delta \si_x=1$ in state $\ketup$, we have to prove that
in re-selection the contribution of the discarded events satisfies $\Mb \si_1\si_2\vert_{disc}\times\mbox{(discard rate)}\rightarrow\half$.
Its analytic form  can be written as
\begin{equation}
\Mb \si_1 \si_2\vert_{disc} = \frac{\bradn\left(\exp(-\quar\sio_\Delta^2/a^2)\sio_c^2\ketup\braup\right)\ketdn}
                                                                {\bradn\left(\exp(-\quar\sio_\Delta^2/a^2)\ketup\braup\right)\ketdn}.
\end{equation}
The denominator yields rate $\sim\half a^{-2}$ of discards, it is vanishing in the WM limit. 
The exponential factor in the numerator can be neglected in the WM limit and we get the following result:
\begin{eqnarray}
\Mb \si_1 \si_2\vert_{disc}&=&2a^2  \bradn\left(\sio_c^2\ketup\braup\right)\ketdn=\\
                                                   &=& 2a^2 \quar    \bradn\{\sio,\{\sio,\ketup\braup\}\}
                                                               \ketdn=a^2.\nonumber
\end{eqnarray}
As we see, the correlation of two subsequent $\sio_x$ polarization WMs diverges on the
discarded events in re-selection. This is in itself a different and stronger anomaly 
than the paradigmatic large but finite mean values obtained in single WMs with post-selection \cite{AAV88}. 
What we  wished to confirm here is that the divergent mean $a^2$ compensates the vanishing rate
$\half a^{-2}$ to yield the finite contribution $\half$ of the discarded outcomes in re-selection.

\emph{Summary, discussion} --- Superoperator formalism has helped us to determine
the  correlation functions of sequential WMs in terms of the quantum expectation values of the
step-wise symmetric  product of the corresponding observables. Condition of interchangeability of WMs within the sequence
has been found. Canonical variables are interchangeable and, without post-selection,
their WM correlation functions coincide with the corresponding correlation functions  of the Wigner function. 
It follows from our result how all $n$-th order correlation functions (moments) of the Wigner function can, in principle, 
be determined directly on the outcome statistics of the sequence of $n$ WMs. This makes sequential WMs 
a tool of direct quantum state tomography (limited normally by the highest available order $n$ in a given experiment). Sequential WMs may
demonstrate quantum paradoxes since the negativity of the Wigner function leads to non-classical statistics of 
sequential WMs, like in ref. \cite{BedBel10}, see also \cite{BedBel11}. Earlier suggestions associated outcomes
of single post-selected WMs with Bohmian velocities \cite{bohmian}. As to the outcomes of sequential WM, our result 
suggests Wigner phase space coordinates as the natural interpretation. (This interpretation proves to be
universal if sequential WM of spin-$\half$ observables is related to Wigner function in Grassmann variables 
introduced in ref.  \cite{CahGlau99}, an issue we leave open here.) Spin-$\half$ observables behave very differently.
Two polarization WMs yield no new information at all compared to single  measurements since the correlation
is determined by the angle between the two polarizers and independent of the quantum state,  
just like for two strong (ideal) polarization measurements. This is more than resemblance. 
We found that  a sequence of $n$ weak or, alternatively,  strong spin-$\half$ measurements 
yield identical $n$-order correlation functions, respectively.

Finally, we studied the marginal case $\ketf=\keti$ of post-selection which we called re-selection 
and found that in sequential WMs it is not equivalent with lack of post-selection. This means that in
sequential WMs with re-selection the discarded statistics matters however close we are to
ideal WMs. This unexpected effect roots in a novel weak value anomaly this time referring to the 
anomalous (divergent) value of the weakly measured (i.e.: in sequential WM) auto-correlation 
on the statistics discarded by re-selection. This phenomenon is a robust feature of sequential
WMs and it is not tractable in terms of standard weak values. As an example, we have 
shown that the correlation of outcomes in double WM of $\sio_x$ in pre-selected state $\keti=\ketup$ 
and post-selected on $\ketf=\ketdn$ will diverge whereas  any correlation larger than $\Vert\si_x\Vert^2=1$
is counter-intuitive.

This work was supported by the Hungarian Scientific Research Fund under Grant No. 103917, and by the EU COST
Action MP1209.

\end{document}